\documentstyle[prl,aps,twocolumn]{revtex}
\begin{document}

%TCIMACRO{
%\TeXButton{twocolumn}{\twocolumn[\hsize\textwidth\columnwidth\hsize\csname @twocolumnfalse\endcsname
%\title{From chaos to disorder: Statistics of the  eigenfunctions of microwave cavities}
%\author{Prabhakar  Pradhan and S. Sridhar}
%\address{Department of Physics \\
%Northeastern University, Boston, Massachusetts 02115}
%\date{\today}
%\maketitle
%
%\begin{abstract}
%We study  the statistics of the experimental  eigenfunctions of chaotic and disordered
%microwave billiards in terms of the moments of their  spatial distributions, such as the 
%Inverse Participation Ratio (IPR)   and density-density auto-correlation. A path from chaos to disorder 
%is described in terms of increasing IPR. In the chaotic, ballistic 
%limit, the data correspond well with universal results from random matrix theory. 
%Deviations from universal distributions 
%are observed due to disorder induced localization, and for the weakly disordered  case the 
%data are well-described by  including  finite conductance and mean free path contributions in the framework
%of nonlinear sigma models of supersymetry. 
%\end{abstract}
%\pacs{05.45.Mt, 05.45.Ac, 03.65.Sq, 84.40.-x}
%\vskip1.5pc]
%}}
%BeginExpansion
\twocolumn[\hsize\textwidth\columnwidth\hsize\csname @twocolumnfalse\endcsname
\title{From chaos to disorder: Statistics of the  eigenfunctions of microwave cavities\\ }
\author{Prabhakar  Pradhan$^{1,2}$  and S. Sridhar$^{1}$ }
\address{$^1$ Department of Physics, Northeastern University, Boston, Massachusetts 02115\\
$^2$Research Laboratory of Electronics, Massachusetts Institute of Technology, Cambridge, MA 02139}
\date{\today}
\maketitle

\begin{abstract}
We study  the statistics of the experimental  eigenfunctions of chaotic and disordered
microwave billiards in terms of the moments of their  spatial distributions, such as the 
inverse participation ratio (IPR)   and density-density auto-correlation. A path from chaos to disorder 
is described in terms of increasing IPR. In the chaotic, ballistic 
limit, the data correspond well with universal results from random matrix theory. 
Deviations from universal distributions 
are observed due to disorder induced localization, and for the weakly disordered  case the 
data are well-described by  including  finite conductance and mean free path contributions in the framework
of nonlinear sigma models of supersymetry.
 
\end{abstract}
\vskip1.5pc]

%
%EndExpansion

\section{\bf INTRODUCTION }

This paper briefly summarizes results and insights gained from expeirmental
studies of eigenfunctions of chaotic and disordered 2D billiards. The
experiments which utilize microwave cavities, enable us to tune the degree
of localization by varying two parameters, the frequency $f$ and the mean
free path $l$. Thus we are able to access the ballistic limit $l\gg \lambda $
(the wavelength) as well as the strongly localized limit $l\prec \lambda $.
In our earlier work \cite{kudrolli95a} we have experimentally shown that the
spatial intensity distribution of an eigenfunction of a chaotic billiard
quantitatively agrees with the Random Matrix Theory (RMT). Recently, we have
obtained further results on the statistics of the eigenfunctions \cite
{pradhan00} of chaotic and disordered systems, which are reviewed in this
paper. The results of our experiments on disordered microwave cavities are
quantitatively described by calculations based upon the supersymmetry sigma
models carried out by Efetov \cite{efetovbook}, Prigodin and Altshuler \cite
{prigodin98}, Mirlin \cite{mirlin00} and others. In this paper we explore
the eigenfunction statistics of chaotic and disordered systems using inverse
participation ratio as a measure of the disorder strength.

\section{\protect\smallskip {\bf Theoretical background: RMT and Nonlinear
Sigma Models }}

It was shown by Berry that the spatial amplitude distribution of the
wave-function of an electron in a chaotic cavity can be generated from the
random superposition of plane waves. This leads to the universal spatial
distribution of the eigenfunction $P(\Psi )$ which is Gaussian, and which
implies that the spatial intensity distribution of the eigen function $%
P(|\Psi |^{2})$ is Porter-Thomas \cite{srednicki}. Spatial correlations are
also obtained, such as the density-density autocorrelation $\left\langle
|\Psi (r)|^{2}|\Psi (r^{^{\prime }})|^{2}\right\rangle
=1+2J_{0}^{2}(k|r-r^{\prime }|)$ for a $2-D$ chaotic billiard.

The moments of the spatial integral of density of an eigenfunction $%
I_{n}=\int |\Psi (\bar{r})|^{2n}d^{3}r$ are important measures of the
localization, particularly the second moment, the Inverse Participation
Ratio $I_{2}$ $(IPR)$, and its statistics $P_{I_{2}}(I_{2})$, which
describes important properties of the chaotic and disordered systems. The
eigenfunctions of a classically chaotic billiard are delocalized wave
functions in the limit of an infinite system. The moments $I_{n}$ have fixed
values with no fluctuations, with $I_{2}=3.0$ in 2-D, i.e. $%
P_{I_{2}}(I_{2})=\delta (I_{2}-3.0)$ \cite{prigodin98,mirlin00}. The universalities
have been studied for some time within the theoretical framework of the RMT.

From the perpective of transport in disordered systems, the chaotic system
can be considered as the ballistic limit corresponding to infinite mean free
path and infinite conductance. Due to the presence of weak disorder, the
universal properties break down, and system properties deviate from the
universal values. A perturbative treatment has been achieved using nonlinear
sigma models of supersymmetry, originally motivated by the problem of
electrons in a disordered metal, which can be termed quantum diffusion \cite
{ekke01}. The supersymmetry approach, which treats an electron in a
potential $V_{imp}(r)$ characterized by a Gaussian distribution of random
disorder due to impurities, enables the calculation of spectral
correlations, such as level spacing statistics, form factor, etc. and
eigenfunction correlations including amplitude $P(\Psi )$ and density $%
P(|\Psi |^{2})$ distributions, and spatial auto-correlations $\left\langle
|\Psi (r)|^{2}|\Psi (r^{^{\prime }})|^{2}\right\rangle $. In the spatially
homogeneous $0$-mode limit of the theory, which corresponds to the ballistic
limit $kl>>1$, the theory reduces to that of the chaotic limit and yields
the results of RMT, such as the Porter-Thomas density distribution,
characteristic of systems with a Gaussian distribution of amplitudes.

With increasing disorder, i.e. with a finite mean free path and a finite
conductance, the system statistics deviate from the RMT, and now the system
can be described by $1D$ sigma models as a perturbative correction with a
finite conductance $g$ and a finite disorder controlled parameter $kl,$
where $k$ is the wave vector and $l$ is the finite mean free path. The
leading term, the 1-mode limit, yields the leading corrections due to
incipient localization, with higher orders (in principle) leading to
increasing localization. The intensity-intensity auto-correlation $%
\left\langle |\Psi (r)|^{2}|\Psi (r^{^{\prime }})|^{2}\right\rangle $ also
can be described by sigma model calculations. For a disordered system,
correlation is large at $r-r^{\prime }=0$ and dies out as with the
increasing distance.

\section{\bf EXPERIMENTAL DETAILS}

Recent developments in experimental techniques have made it possible to
measure the eigenvalue and eigenfunction properties of thin cylindrical
microwave cavities - these studies have enabled quantitative studies of
issues in Quantum Chaos and Localized media on laboratory length scales with
very high precision experiments.

\smallskip In $2D$, Schr\"{o}dinger and Maxwell equation map onto each other
in the sense that the Helmholtz equation for the $z$-component of the
electric field becomes $(\nabla ^{2}\,+\,k^{2})\psi =0$, where $\psi =E_{z}$%
. $2D$ cavities were made between two parallel copper plates, and the shape
and size of a cavity was made according to the need. The height of the
cavity is $d=6$ mm, and the cavity is effectively $2D$ for the range of
microwave wave lengths used in the experiments. The electric field
distribution for the Transverse Magnetic (TM) modes of such a cavity obeys
the Schr\"{o}dinger wave equation in two dimensions \cite
{sridhar91,kudrolli94,microwavechaos}.

For chaotic billiard experiments, the shape of the cavity was made like a
Sinai Stadium or like Sinai Billiards. For disordered closed billiard
experiments, the cavity has closed rectangular ( $44cm$ x $21.8cm$ )
boundaries, and the disordered scattering centers were imposed by placing
randomly circular $2D$ plates of $1cm$ radius. The scattering mean free path
was changed by changing the number of the scattering centers. Several
realizations of a scattering system were done by changing the scattering
positions with a random number generator and keeping the total number of
scatterers as same. A variety of experiments on closed and open quantum
chaotic systems have been carried out using such microwave cavities\cite
{sridhar91,kudrolli94,microwavechaos}. A particularly powerful aspect of the
experiments is the ability to directly measure wavefunctions using cavity
perturbation techniques \cite{sridhar91}.

Experiments on the eigenvalue statistics have enabled stringent tests of the
eigenvalue properties of chaotic systems \cite{kudrolli94}. Experiments on
disordered billiards show a very good realization of 2D disordered systems%
\cite{kudrolli95a}, and which motivated further theoretical works in this
field \cite{falko94,prigodin98,mirlin00}. Most recently, we have carried out
a systematic study of the statistics of the experimental eigenfunctions for
chaotic and disordered systems\cite{pradhan00}. These papers show tests of
several analytical results of RMT and the nonlinear model of supersymmetry
which was developed mainly for disordered granular media and mesoscopic
systems. At present, the eigenfunctions of mesoscopic systems are difficult
to access for controlled experiments due to their smaller size, and
consequently microwave cavity experiments play an unique and important role
for wave function statistics.

\section{\bf EXPERIMENTAL\ EIGENFUNCTIONS AND THEIR STATISTICS }

The utility of analyzing eigenfunctions in terms of their $IPR$ is discussed
here for a few representative eigenfunctions. For convenience, we will use
the notation $I_{2}{\bf =}\int |\Psi (r)|^{4}dv=\int u^{2}dv$, $u=|\Psi
(r)|^{2}$, $w=(I_{2}-3)/6$, and $I_{1}{\bf =}\int |\Psi (r)|^{2}dv=1$, where
the integration is done over the $2D$ volume. More than $250$ eigenfunctions
are analyzed for chaotic and disordered systems, and each eigenfunction has $%
3200$ spatial data points.

The representative eigenfunctions of the disordered billiard with $N=36$
scatterers have been shown in Figs.1(a), (b), (c) and (d). Fig.1(a) shows
the eigenfunction of a disordered billiard with a low eigenfrequency $%
f=3.84GHz$ has localized eigenstate  $IPR$ with value $I_{2}=11.22.$ With
increased frequency (energy), the same disordered medium shows more
delocalized eigenstates with decreasing $IPR$ values : Fig.1(b) $I_{2}=9.64$
at $4.49GHz$, Fig.1(c) $I_{2}=7.52$ at $6.37GHz$, and Fig.1(d) $IPR$ value $%
I_{2}=4.36$ at $7.20GHz$. This can be regarded as a tuning the degree of
localization by changing the frequency (energy). In fact, we have shown \cite
{pradhan00} that this tuning path follows a power law decay with an exponent 
$\frac{1}{2},$ consistent with theoretical calculations \cite{fyodorov97}.

Fig.1(e) and (f) show representative eigenfunctions of a chaotic billiard,
which are ``delocalized'', $IPR$ $I_{2}=3.13$ at $6.05GHz$ (Fig.1(e)), and $%
I_{2}=3.02$ at $7.54GHz$ (Fig.1(f)). These eigenfunctions have their $IPR$
values close to the universal value $\langle I_{2}\rangle =3$ in $2D$.

A comparison of experimental results of chaotic and localized eigenfunction
statistics are shown in Figure 2 for (i) spatial density distribution of
eigenfunctions, (ii) statistics of eigenfunctions interms of their IPR
values, and (iii) intensity auto-correlation of\ the\ eigenfunctions.

\subsection{\bf Spatial density distribution of eigenfunctions}

Figure 2(a) shows that the spatial intensity distribution $P(|\Psi |^{2})$
of eigenfunctions of a chaotic system obeys the Porter-Thomas distribution.
The P-T distribution is obtained from RMT, as well as random superposition
of plane waves and the 0-D sigma model. The deviation from the PT
distribution due to finite localization of a disordered medium is shown in
Figure 2(b). This can be modeled by introducing finite mean free path in the
problem and a quantitative description of the data has been achieved \cite
{falko94}.

\subsection{\bf Statistics of IPR of eigenfunctions }

The distribution $P_{I_{2}}(I_{2})$ of the $IPR$ $I_{2}$for the chaotic
billiards is shown in the Fig.2(c). The distribution is nearly symmetric
around $\left\langle I_{2}\right\rangle =3.$ RMT calculations indicate that
there should be no fluctuations, but this only applies for an infinite
system. The finite width of the distribution is due to the boundary
scattering, and can be quantitatively described by large but finite $2kl$
and $g$ values \cite{muzy95}. A quantitative description is given later.

In contrast the $IPR$ distribution (Fig.2(d)) for a disordered billiard with
mean free path $l=5.1cm$ ($N=36$ scatterers) is strongly asymmetric and has
a mean value much larger than the universal value of $3.0$. These data can
be modeled with a finite conductance $g$ . A dimensionless conductance can
be defined as $g=\ln (R/l)/\langle w\rangle $, where $R$ is the system size
and $l$ is the mean free path and $\langle ..\rangle $ is the realization
average for a fixed ``disordered strength'' $2kl.$

Recent theoretical calculations by Prigodin and Altschuler \cite{prigodin98}
have shown that when $I_{2}\gg \langle I_{2}\rangle $ , the IPR distribution
follows an exponential decay law

\begin{equation}
P_{I_{2}}(I_{2})=C_{2}\sqrt{\frac{g}{I_{2}}}\exp (-\frac{\pi }{6}gI_{2})
\end{equation}

while $P(I_{2})$ for $I_{2}<\langle I_{2}\rangle $ is \cite{prigodin98} 
\begin{equation}
P_{I_{2}}(I_{2})=C_{1}\frac{g}{2}\exp [-\frac{g}{6}(I_{2}-\langle
I_{2}\rangle )-\frac{\pi }{2}e^{-\frac{g}{3}(I_{2}-\langle I_{2}\rangle )}],
\end{equation}
\smallskip where $C_{1}$ and $C_{2}$ are the normalization constants. For
the sample with finite mean free path $l=5.1cm,$ and assuming random phase
approximation within a small window where $2kl$ is fixed and independent of $%
IPR$ distribution \cite{pradhan97}, experimental data matches well with Eq.1
with a conductance value $g=1.0$ as shown in Fig.2(d). The distribution is
also asymmetric beyond the universal average value $\left\langle
I_{2}\right\rangle =3,$ for the chaotic billiards.

For the chaotic billiard, the data matches well with Eq.2, as shown in
Fig.2(c) for the finite conductance $g=7.8$ and $2kl=37,$ quite large values
that we can consider the system as infinite, i.e., the length scale same as
the system size. Values of $g$ are consistent with the parameters of the
experimental microwave cavity.

\subsection{\bf Intensity-intensity auto-correlation of\ the\ eigenfunctions}

In earlier work we have shown that the intensity-intensity auto-correlations
for the chaotic cavity has a universal behavior associated with Friedel
oscillations \cite{prigodin95}. We have extended the density
auto-correlation calculation from chaotic to localized eigenfunctions, and
analyzed the results analytically and numerically. The intensity-intensity
auto-correlation for an arbitrary disordered strength $2kl$ can be
calculated as follows. Let us define $K(r)=|%
%TCIMACRO{\func{Im}}
%BeginExpansion
\mathop{\rm Im}%
%EndExpansion
G(r^{\prime })|^{2}/(\pi \nu )^{2}$ , where $G(|r-r^{^{\prime
}}|)=\left\langle r|(E-H)^{-1}|r^{\prime }\right\rangle $ is the Green
function of the disordered system Hamiltonian, then one can show that \cite
{prigodin94}

$K(r)=|\frac{1}{\pi }\int_{-\infty }^{\infty }\frac{1}{1+y^{2}}J_{0}[kr(1+%
\frac{1}{2kl}y)]dy|^{2}$.

where $J_{0}$ is the zeroth order Bessel function.

For chaotic billiards the auto-correlation follows an analytical form

\begin{equation}
\left\langle \Psi ^{2}(r)\Psi ^{2}(r^{\prime })\right\rangle
=1+(I_{2}-1)J_{0}^{2}(r-r^{\prime })
\end{equation}

This also corresponds to that for a flat disordered potential with Gaussian
fluctuations, when $2kl>>1$. We extend the previous calculations for the
finite localization case. Repeating the earlier calculations of \cite
{prigodin95} with a finite $kl$ value, it can be easily shown that, the
expression for the auto correlation can be approximately expressed in a
region $r-r^{\prime }\precsim l$ by a decay length scale of scattering mean
free path length $l$ as follows. 
\begin{eqnarray}
\left\langle \Psi ^{2}(r)\Psi ^{2}(r^{\prime })\right\rangle &\simeq
&1+(I_{2}-1)K(k|r-r^{\prime }|) \\
&\simeq &1+(I_{2}-1)J_{0}^{2}(k|r-r^{\prime }|)e^{-\frac{k|r-r^{^{\prime }}|%
}{kl}}
\end{eqnarray}

\smallskip With $kl$ large, the above expression, Eq.4 and 5, will converge
to the expression of a chaotic system, Eq.3.

We have plotted the value of $\left\langle \Psi ^{2}(r)\Psi ^{2}(r^{\prime
})\right\rangle $ for different types of chaotic billiards in Fig. 2(e) and
it agrees well with Eq.3.

For the disordered billiards, $K$ was solved numerically, and also
approximate analytical results have been plotted in Fig. 2(f), and were
compared with the results of $\left\langle \Psi ^{2}(r)\Psi ^{2}(r^{\prime
})\right\rangle $ calculated from experimental data. The agreement is
quantitatively excellent in the described region. For Sinai billiards, $%
2kl=37$ is a very large number and described well for a Gaussian
distribution of the wave functions (Fig. 2(f)). For the disordered
billiards, the matchings are shown for the $2kl=7$ and $3$ (Fig. 2(f)). The
correlation is large at $r-r^{\prime }=0,$ i.e. equals $\left\langle \Psi
^{4}(r)\right\rangle =I_{2}$, and the correlation decays to zero with
increasing spatial distance, as the auto-correlation is negligible at a
larger distance beyond the scale of the localization length.

\section{\protect\smallskip {\bf CONCLUSIONS}}

In this paper we have reviewed statistical properties of wave functions of
chaotic and disordered systems using the second moment of the spatially
integrated intensity of the eigenfunctions. We have shown that the
statistics agrees well with random matrix theory for chaotic systems, and
with the leading perturbative correction of the nonlinear sigma model of
super symmetry for incipient disordered media. The perspective that emerges
from the microwave experiments is summarized in Figure 2 in terms of spatial
intensity distribution (Fig.2(a,b)), statistics of Inverse Participation
Ratio ($IPR$) of eigenfunctions (Fig.2(c,d)) , and intensity-intensity
auto-correlation function (Fig.2(e,f)). The universal ($2D$) results of
Quantum Chaos, characterized by Gaussian fluctuations of $\Psi $, are
achieved in chaotic billiards, albeit with small corrections due to
correlations induced by boundary scattering. The importance of correlations
due to coherent interference between scattering events are amplified in the
model disordered billiards. By tuning $kl$ we can observe a path from
extended ($I_{2}\sim 4$) states to strongly localized ($I_{2}\sim 20$)
states, as can be seen in a $I_{2}$ $vs.$ $f^{2}$ plot (Fig.2 in Ref. \cite
{pradhan00}). This is similar to tuning from a metal to insulator by
increasing energy! This is also accompanied by deviations from the
Porter-Thomas ($PT$) distributions, and strong level-to-level fluctuations,
leading to a strongly asymmetric distribution $P_{I_{2}}(I_{2}),$ and strong
spatial decay of correlations, all indicating non-Gaussian amplitude
distributions. For not too strongly localized states, we have shown
agreement with the supersymmetry theories, parameterized by finite
conductivity $g$ (Figures: 2(b), 2(d) and 2(f) ). The conductivity $g$
depends monotonically on $kl$, with the ballistic chaotic limit
corresponding to $g,kl\rightarrow \infty $ . A systematic study of wave
function statistics is under way to explore more strongly localized states
for which however there is no available theory currently.

\smallskip The eigenvalue and eigenfunction statistics of a disordered
medium for finite $g$ have been calculated \cite{prigodin98} by solving a
Perron-Frobenius operator of a classical diffusion equation, whose (real)
eigenvalues are related to the poles of a classical zeta function. Recently,
a direct experimental observation of the complex poles of the Liouville
operator for an open chaotic system, the so-called classical Ruelle-Pollicot
resonances, have been observed by us in microwave experiments on open n-disk
systems \cite{pance00,lu99,lu00,gaspard93}. This latter work thus provides a physical
reality to these eigenvalues in open systems which have played a significant
role in the theory of the disordered systems, and provides a close
connection between classical and quantum diffusion.

While the nonlinear sigma model provides a useful, even quantitative,
framework to discuss arising from coherent interference leading to
localization, its perturbative approach limits it to the leading corrections
to the ballistic or RMT\ limit. The experiments are not limited to the
regime of large $g$ i.e. $kl\gg 1$, but can easily reach the strong
localization limit $kl\prec 1$. This regime requires higher order
calculations beyond the $1$-mode that are quite formidable. Other approaches
to this problem deserve attention. One possibility that intrigues us is the
information theoretic approach based upon maximum entropy employed by Kumar
and collaborators \cite{maxentropy}. While this has thus far been applied
only to quasi $1D$, extensions to higher dimensions would be worth pursuing.

It is a pleasure to dedicate this paper to Professor Narendra Kumar. PP had
the honor of carrying out his Ph.D. thesis research under Professor Kumar's
guidance. SS has had the pleasure of numerous discussions on physics that
have always been enthralling and memorable. This work was supported by
NSF-PHY-0098801.

%TCIMACRO{\TeXButton{narrowtext}{\narrowtext}}
%BeginExpansion
\narrowtext%
%EndExpansion

%TCIMACRO{
%\TeXButton{figure1}{\begin{figure}
%\caption{ (a,b,c,d).  Representative eigenfunctions  of  a disordered billiard with number of 
%scatterers is  36  :  (a)  $IPR$ value $I_{2}=11.22 $ at $f=3.84GHz$,
%(b) $I_{2}=9.64$ at $4.49GHz$,  
%(c) $I_{2}=7.52$ at $6.37GHz$ and 
%(d) $I_{2}=4.36$ at $7.20GHz$.
%The plots show  that  the eigenfunctions are more localized ( larger IPR value) at a lower 
%frequency and more delocalized (smaller IPR value) at a higher  frequency.\\
%Fig.1(e,f)  Representative eigenfunctions of a chaotic billiard: 
%(e) IPR value  $I_{2}=3.13$ at $6.05GHz$ , and  (f)  IPR value  $I_{2}=3.02$ at 
%$7.54GHz$.  See the text for details. }
%\label{fig1}
%\end{figure}}}
%BeginExpansion
\begin{figure}
\caption{(a,b,c,d) Representative eigenfunctions  of  a 
disordered billiard with number of 
scatterers is  36  :  (a)  $IPR$ value $I_{2}=11.22 $ at $f=3.84GHz$,
(b) $I_{2}=9.64$ at $4.49GHz$,  
(c) $I_{2}=7.52$ at $6.37GHz$ and 
(d) $I_{2}=4.36$ at $7.20GHz$.
The plots show  that  the eigenfunctions are more localized ( larger IPR value) at a lower 
frequency and more delocalized (smaller IPR value) at a higher  frequency.\\
Fig.1(e,f)  Representative eigenfunctions of a chaotic billiard: 
(e)  $I_{2}=3.13$ at $6.05GHz$ , and  (f)    $I_{2}=3.02$ at 
$7.54GHz$.  }
\label{fig1}
\end{figure}%
%EndExpansion

%TCIMACRO{
%\TeXButton{figure2}{\begin{figure}
%\caption{ (a,b) . The spatial intensity distribution of experimental eigenfunctions: 
% 2(a) Porter-Thomas distribution for  chaotic cavity,
% and 2(b) non  Porter-Thomas distribution for  disordered cavity.\\
%Fig. 2(c,d).   2(c) The IPR  $P_{I_2} (I_2 )$  distribution of  the  chaotic Sinai-stadium 
%billiard   is nearly symmetric around  the universal  mean value $3.0$ . 2(d)  
%The  IPR distribution $P_{I_2} (I_2 )$ of the disordered billiards  is strongly  
%asymmetric and non-Gaussian. The lines 
%represent  calculations based on the nonlinear sigma-models.\\
%Fig 2(e,f). Intensity auto-correlation $\left\langle \Psi ^{2}(r)\Psi ^{2}(r^{\prime
%})\right\rangle$ of  eigenfunctions of a   2(e) Sinai-stadium  billiards, and 
%2(f) disordered billiards  with different fixed disordered strengths 
%$2kl$, experiment ( dotted lines),  Eq.$(4)$ (dashed lines), 
%and numerical solution of Eq.$(4)$  with $2kl$ =$13$  and $7$ (solid lines).
%}
%\label{fig2}
%\end{figure}}}
%BeginExpansion
\begin{figure}
\caption{ (a,b) . The spatial intensity distribution of experimental eigenfunctions: 
 2(a) Porter-Thomas distribution for  chaotic cavity,
 and 2(b) non  Porter-Thomas distribution for  disordered cavity.\\
Fig. 2(c,d).   2(c) The IPR  $P_{I_2} (I_2 )$  distribution of  the  chaotic Sinai-stadium 
billiard   is nearly symmetric around  the universal  mean value $3.0$ . 2(d)  
The  IPR distribution $P_{I_2} (I_2 )$ of the disordered billiards  is strongly  
asymmetric and non-Gaussian. The lines 
represent  calculations based on the nonlinear sigma-models.\\
Fig 2(e,f). Intensity auto-correlation $\left\langle \Psi ^{2}(r)\Psi ^{2}(r^{\prime
})\right\rangle$ of  eigenfunctions of a   2(e) Sinai-stadium  billiards, and 
2(f) disordered billiards  with different fixed disordered strengths 
$2kl$, experiment ( dotted lines),  Eq.$(4)$ (dashed lines), 
and numerical solution of Eq.$(4)$  with $2kl$ =$13$  and $7$ (solid lines).
}
\label{fig2}
\end{figure}%
%EndExpansion

\end{document}